\documentclass[11pt,letterpaper]{article}

\usepackage[margin=1in]{geometry}
\usepackage[utf8]{inputenc}
\usepackage[T1]{fontenc}
\usepackage{times}
\usepackage{amssymb}

\usepackage{amsmath,amsfonts,bm}

\def\eqref#1{equation~\ref{#1}}

\def\1{\bm{1}}

\DeclareMathAlphabet{\mathsfit}{\encodingdefault}{\sfdefault}{m}{sl}
\SetMathAlphabet{\mathsfit}{bold}{\encodingdefault}{\sfdefault}{bx}{n}

\usepackage{nicefrac}
\usepackage{microtype}
\usepackage{xcolor}
\usepackage{graphicx}
\usepackage{float}
\usepackage{algorithm}
\usepackage{algpseudocode}
\usepackage{booktabs}
\usepackage{multirow}
\usepackage{makecell}
\usepackage[font=small,labelfont=bf]{caption}
\usepackage[authoryear,round]{natbib}
\setcitestyle{citesep={;},aysep={,},yysep={;}}
\usepackage{xurl}
\usepackage{fancyhdr}
\usepackage[hidelinks,bookmarksnumbered]{hyperref}

\fancypagestyle{preprint}{%
  \fancyhf{}
  \fancyhead[L]{\normalfont\normalsize Preprint}
  \fancyfoot[C]{\thepage}

}
\newcommand{\papertitle}{RADNPO: Reference-free Adaptive Negative Preference Optimization for LLM Unlearning}
\newcommand{\paperauthors}{%
  Shenghan Tan\textsuperscript{1}\quad
  Ziyi Zhou\textsuperscript{1}\quad
  Wenpeng Hu\textsuperscript{2}\quad
  Mengyuan Zhang\textsuperscript{1}\thanks{Corresponding author.}\\[0.5em]
  \normalsize\textsuperscript{1}Beihang University\qquad
  \normalsize\textsuperscript{2}Peking University
}
\newcommand{\paperauthornames}{Shenghan Tan, Ziyi Zhou, Wenpeng Hu, Mengyuan Zhang}

\hypersetup{
  pdftitle={\papertitle},
  pdfauthor={\paperauthornames},
  pdfsubject={Machine unlearning for large language models},
  pdfkeywords={machine unlearning, large language models, preference optimization, RADNPO}
}
\title{\LARGE\bfseries\papertitle}
\author{\paperauthors}
\date{}

\begin{document}
\maketitle
\thispagestyle{preprint}

\begin{abstract}
Large language models (LLMs) can memorize sensitive, private, or copyrighted content during pre-training, making machine unlearning necessary for removing targeted knowledge. Recent preference optimization (PO)-based unlearning methods improve stability over gradient ascent (GA)-based methods by introducing alignment-style objectives, which effectively suppress the probability of forget targets.
However, target suppression alone does not sufficiently constrain the next-token distribution after unlearning. Existing methods provide limited control over how suppressed probability mass is redistributed and insufficiently adapt forgetting strength to target confidence and distributional concentration.  Even after target suppression, probability mass may remain concentrated on a few non-target tokens, potentially producing repetitive or uninformative outputs.
To address these limitations, we propose \textbf{R}eference-free \textbf{AD}aptive \textbf{N}egative \textbf{P}reference \textbf{O}ptimization (\textbf{RADNPO}), which explicitly guides next-token probability redistribution. Specifically, RADNPO contrasts each forget target with alternative tokens favored by the current next-token distribution and adaptively modulates token-level forgetting strength using target confidence and next-token concentration.
Experiments on TOFU and MUSE demonstrate that RADNPO achieves a better trade-off between forgetting quality and model utility than current baselines.
\end{abstract}

 \section{Introduction}

LLMs derive strong general capabilities from large-scale pre-training \citep{carlini2021extracting, touvron2023llama}, and also memorize sensitive, private, and copyrighted content during this process. The memorized content can lead to  privacy leakage \citep{das2025security}, toxic generation \citep{jang2023knowledge}, and unauthorized reproduction \citep{huang2024position}, introducing safety and legal risks.
Since retraining from scratch is prohibitively expensive, machine unlearning (MU) \citep{liu2025rethinking, nguyen2025survey} become a practical way to remove the influence of unwanted data from pretrained LLMs.

Early LLM unlearning methods often build on GA-based \citep{maini2024tofu} forgetting methods, which is intuitive but unstable and can lead to catastrophic forgetting \citep{zhang2024negative}. PO-based \citep{rafailov2023direct, zhang2024negative} methods mitigate this instability by replacing vanilla GA with a relative preference objective. 
Negative preference objectives such as NPO \citep{zhang2024negative} and SimNPO \citep{meng2024simpo} suppress forget responses without constructing explicit positive alternatives for the corresponding prompts. However, reducing the forget target probability alone does not specify how probability mass should be distributed among alternative
tokens.
AltPO \citep{mekala2025alternate} provides additional guidance by pairing forget responses with generated alternative answers. 
This motivates our approach to explicitly contrast each forget target with high-probability non-target alternatives selected from the model's current next-token distribution.

 Current PO-based unlearning methods \citep{zhang2024negative,fan2024simplicity,mekala2025alternate} constrain only a limited aspect of the token-level output distribution: they require the forget target probability to be reduced, but do not control how the suppressed probability mass is redistributed among non-target tokens. Without explicit guidance, the model may leave the original target dominated state but redistribute probability mass in an uncontrolled way, shifting preference toward implausible, repetitive, or uninformative alternatives. 
Therefore, target suppression alone leaves the post-unlearning distribution underspecified, potentially leading to degenerate generation behavior.

\begin{figure}[H]

  \centering
  
  \includegraphics[width=\linewidth]{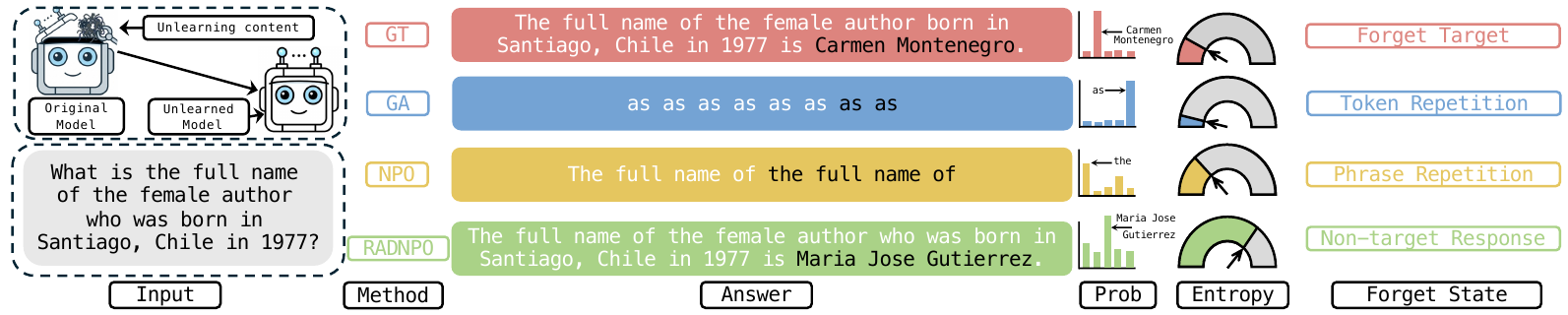}
  \caption{\textbf{Different unlearning methods and their corresponding generation behaviors.} GT is ground truth. GA severely degrades general generation. NPO lowers the probability of forget target. In contrast, RADNPO achieves a less concentrated post-unlearning distribution.}
  \label{fig:fig1}

\end{figure}

Beyond the choice of alternatives, the local optimization state also matters. Some forget targets remain highly probable under the current model and still require substantial suppression, whereas others have already been suppressed but reside in highly concentrated local distributions. These states cannot be fully characterized by the target probability alone.
Existing sequence level NPO objective  primarily adapt to target likelihood and do not explicitly account for target persistence and post-unlearning distributional concentration jointly. Consequently, as the target likelihood decreases, the forgetting strength may weaken even when the local distribution remains highly concentrated. 
This mismatch can leave probability mass concentrated on a small subset of non-target tokens despite target suppression. This regime, which we term low-entropy forgetting, may produce repetitive or uninformative outputs. The desired semantic response after unlearning varies across applications. Our focus is therefore on guiding next-token probability redistribution, rather than prescribing a specific response.
Figure~\ref{fig:fig1} illustrates this distinction.

To address this, we propose \textbf{R}eference-free \textbf{AD}aptive \textbf{N}egative \textbf{P}reference \textbf{O}ptimization (\textbf{RADNPO}). RADNPO combines two complementary mechanisms. 
First, a local contrastive objective compares each forget target with alternative tokens favored by the current next-token distribution, providing an explicit redistribution signal. Second, an adaptive scaling mechanism adjusts the token-wise objective coefficient based on the model's confidence in the forget target, which we refer to as target confidence, and the concentration of the next-token distribution, characterized by the next-token partial entropy.
Together, these designs allow RADNPO to formulate unlearning as state-aware reshaping of local next-token distributions rather than target suppression alone.

In summary, our contributions are as follows:

\textbullet~  We highlight an under-examined failure mode of PO-based unlearning:  target suppression can coexist with an excessively concentrated post-unlearning distribution and degraded generation.

\textbullet~ We propose 
\textbf{RADNPO}, a reference-free adaptive negative preference optimization method that combines a contrastive log-odds objective with an adaptive coefficient scaling mechanism. The former explicitly guides next-token probability redistribution, while the scaling mechanism modulates forgetting strength for each token based on target confidence and distributional concentration.

\textbullet~ Extensive experiments on TOFU \citep{maini2024tofu} and MUSE \citep{shi2024muse} demonstrate that RADNPO improves the trade-off between forgetting quality and model utility over baselines  while preserving retained knowledge.

\section{Related Work}

\textbf{Machine unlearning. } 
Machine Unlearning (MU) aims to remove the residual influence of specific training data from parameterized models, serving as a technical solution to the "right to be forgotten" \citep{sakaguchi2021winogrande}. Initially stemming from the context of image classification in computer vision \citep{golatkar2020eternal,jia2023model}, MU has rapidly expanded its scope to encompass diverse domains, including text classification \citep{jiang2025backdoor, fan2024challenging}, text-to-image generation \citep{kurmanji2023towards}, and federated learning \citep{liu2024survey, halimi2022federated}. MU approaches are broadly divided into two categories: exact unlearning and approximate unlearning. Exact unlearning is achieved by retraining the model from scratch on the retain set, a process universally acknowledged as the gold standard \citep{bourtoule2021machine}. In contrast, approximate unlearning avoids the prohibitive overhead of full retraining, achieving targeted knowledge erasure through efficient fine-tuning or localized model editing \citep{dong2025machine}.

\textbf{LLM unlearning. }
With the rapid advancement of foundational LLMs, MU has emerged as a crucial paradigm to selectively eliminate undesirable data influences while strictly preserving the model utility.  Existing methods include base model fine-tuning \citep{jang2023knowledge}, in-context learning \citep{pawelczyk2023context}, model editing \citep{patil2023can}, and preference optimization \citep{zhang2024negative}. 
Benchmarks such as TOFU  and MUSE have shown that current methods often struggle to balance forgetting quality and model utility. Recent work increasingly adopts preference optimization \citep{zhang2024negative, fan2024simplicity} due to its better optimization stability, but these methods are still largely designed around suppressing the target response.  
This leaves the post-unlearning local distribution under constrained, particularly with respect to how probability mass is redistributed after target suppression, which motivates our work.

\textbf{Preference optimization. }
Motivated by the critical need to align LLMs with human preferences, reinforcement learning from human feedback (RLHF) \citep{christiano2017deep} was introduced. This framework typically achieves such alignment through Proximal Policy Optimization \citep{schulman2017proximal} paired with a separate reward model. Direct Preference Optimization (DPO) \citep{rafailov2023direct} bypasses the reward model to optimize the policy directly on preference data. However, it still necessitates a frozen reference model for KL divergence computation, incurring substantial memory overhead. To alleviate this problem, subsequent studies streamline the optimization process from different angles. For example, KTO \citep{ethayarajh2024kto} relaxes the reliance on paired preference data, while SimPO \citep{meng2024simpo} explicitly discards the reference model by employing a length-normalized implicit reward. Notably, Odds Ratio Preference Optimization (ORPO) \citep{hong2024orpo} presents a highly efficient, single-stage solution. By introducing an odds ratio-based penalty term, it  merges SFT and preference alignment, obviating the need for a reference model.

\section{Preliminary}

\subsection{Problem formulation of LLM unlearning} The pre-training dataset $\mathcal{D}$ comprises a forget set $\mathcal{D}_f$ and a retain set $\mathcal{D}_r$. Given an LLM parameterized by $\theta$, the goal of LLM unlearning is to remove the influence of forget targets in $\mathcal{D}_f$, while preserving the model's general capabilities on $\mathcal{D}_r$. This dual objective can be formulated as the following optimization problem \citep{jin2025unlearning, dorna2025openunlearning}:
\begin{equation}
\label{eq:preliminary}
\min_{\theta} \mathcal{L}_f(\theta; \mathcal{D}_f) + \mathcal{L}_r(\theta; \mathcal{D}_r),
\end{equation}

where $\mathcal{L}_f$ and $\mathcal{L}_r$ denote the forget and retain losses, respectively.
Existing methods commonly adopt GA-based unlearning  \citep{thudi2022unrolling} to reduce the likelihood of forget targets in $\mathcal{D}_f$ by minimizing their log-likelihood, i.e., $\mathcal{L}_f=\mathbb{E}_{(x,y) \in \mathcal{D}_{f}}\left[\log \pi_\theta(y|x)\right]$, with $\pi_\theta$ denoting the output distribution of the unlearning model. 
The retain loss $\mathcal{L}_r$ is typically instantiated as the standard next-token prediction or Kullback--Leibler (KL) regularization to preserve model utility on $\mathcal{D}_r$. 

\subsection{LLM unlearning with preference optimization}
To mitigate the instability of GA-based forgetting, recent methods introduce preference optimization into LLM unlearning. Notably, NPO \citep{zhang2024negative} bridges the LLM alignment and machine unlearning by repurposing the framework DPO \citep{rafailov2023direct}, while requiring only negative samples from the forget set. 
Formally, the forget loss $\mathcal{L}_{\text{NPO}, \beta}(\theta)$ and its gradient in NPO are given by:

\begin{align}
\label{eq:npo_loss}
\mathcal{L}_{\text{NPO}, \beta}(\theta) &= \frac{2}{\beta} \mathbb{E}_{(x,y) \in \mathcal{D}_{f}} \left[ \log  \left( 1 + \left( \frac{\pi_\theta(y|x)}{\pi_{\text{ref}}(y|x)} \right)^\beta \right) \right], \\
\label{eq:npo_gradient}
\nabla_{\theta}\mathcal{L}_{\text{NPO}, \beta}(\theta) &= \mathbb{E}_{(x,y) \in \mathcal{D}_{f}} \left[ \underbrace{\frac{2\pi_\theta(y|x)^{\beta}}{\pi_\theta(y|x)^{\beta}+\pi_{\text{ref}}(y|x)^{\beta}} }_{\mathrm{adaptive~weight}} \underbrace{\nabla_{\theta} \log \pi_{\theta}(y \mid x)}_{\mathrm{GA~gradient}} \right],
\end{align}

where $\beta > 0$ is the temperature parameter,  $\pi_{ref}$ denotes the reference model distribution prior to unlearning. 
As shown in (\ref{eq:npo_gradient}), 
NPO introduces an adaptive weight into the standard GA gradient through the reference model. This weight dynamically modulates the forgetting strength according to the relative likelihood of the forget target under $\pi_{\theta}$ and $\pi_{ref}$, thereby stabilizing GA-based forgetting and preventing catastrophic utility degradation ( Details in Appendix \ref{sec:appendix_npo_grad} ).

\section{RADNPO: Method and Analysis}
\label{sec:method}

\begin{figure}[H]
  \centering
  \includegraphics[height=6.5cm]{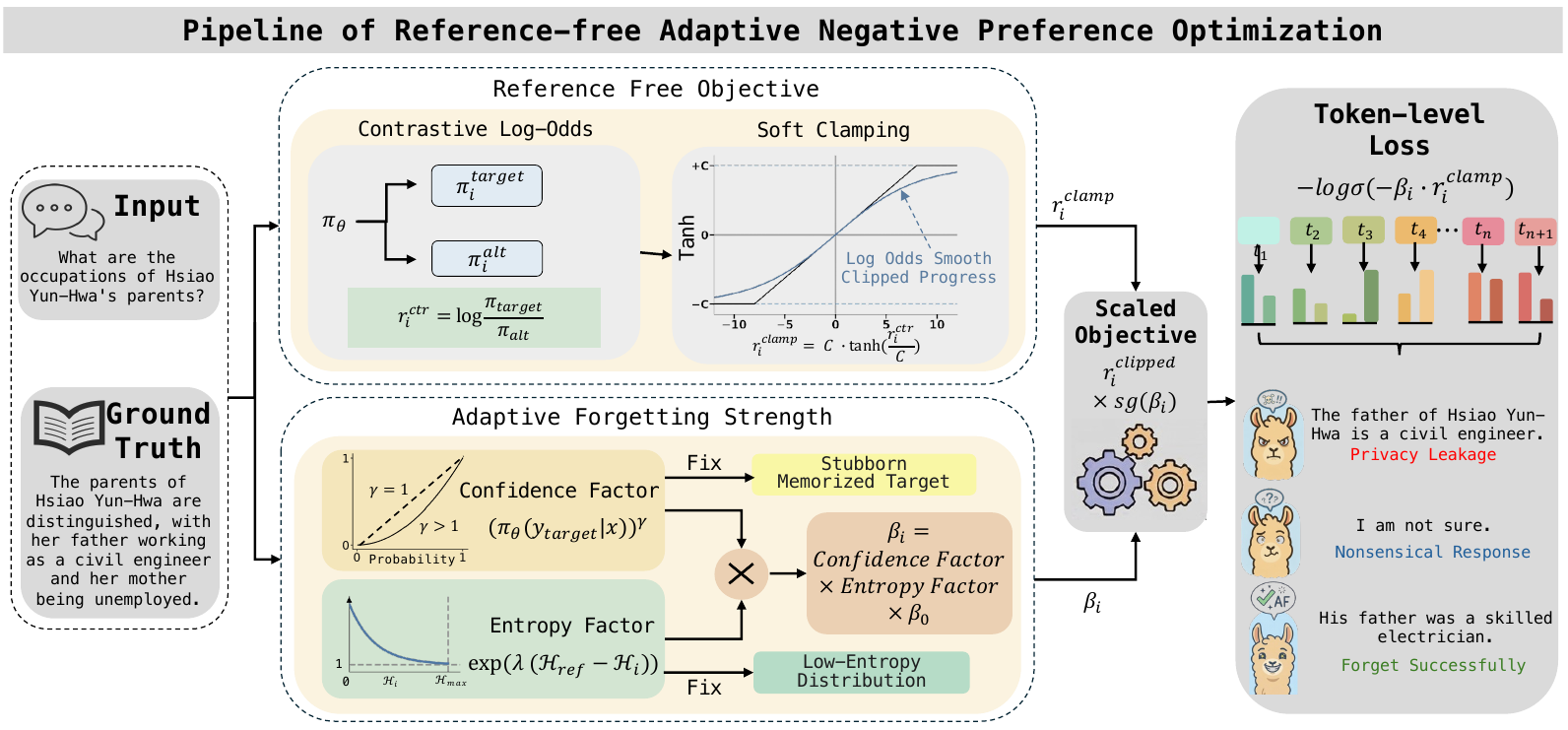} 
  \caption{\textbf{Overview of method:} RADNPO contrasts each forget target with high-probability non-target candidates under the current next-token distribution, applies soft clamping for stable optimization, and adaptively modulate the forgetting strength using target confidence and next-token entropy. The confidence factor strengthens forgetting for high-confidence forget targets, while the entropy factor increases the relative update scale in low-entropy regimes.}
  \label{fig:main}
\end{figure}

Existing PO-based unlearning methods primarily optimize target suppression, while providing limited control over the resulting next-token distribution. Moreover, appropriate forgetting strength depends on the token-level state: some forget targets remain highly probable, whereas others have already been suppressed but remain in highly concentrated local distributions. To address these limitations, we propose Reference-free Adaptive Negative Preference Optimization (RADNPO), which combines a reference-free contrastive objective with a token-level adaptive scaling mechanism. The contrastive objective compares each forget target with high probability non-target candidates to guide the redistribution, while the adaptive mechanism modulates forgetting strength according to target confidence and distributional concentration, as illustrated in Figure~\ref{fig:main}.

\subsection{Reference-Free Contrastive Log-Odds Objective}
\label{para:4.1}

To guide local probability redistribution, we compare each forget target with alternative tokens selected from the current next-token distribution. This comparison introduces a local contrastive objective without relying on a reference model. We further apply soft clamping to limit the influence of extreme contrastive margins.

\paragraph{Contrastive log-odds.}
Inspired by the odds-ratio formulation in ORPO \citep{hong2024orpo}, we construct the forget objective by directly contrasting each forget target with non-target alternatives. 
Rather than using the probability mass of the entire remaining vocabulary \citep{yang2026catnip}, we restrict the comparison to a local set of alternative tokens.
Specifically, for a forget target $y_i$ with context $(x,y_{<i})$, let $\mathcal{V}_{i,K}^{\mathrm{alt}}$ 
denote the set of $K$ highest-probability tokens excluding $y_i$.
We define the target probability and the aggregate probability of these alternatives as
\begin{equation}
\pi_i^{\mathrm{target}}
=
\pi_\theta(y_i \mid x,y_{<i}),
\qquad
\pi_i^{\mathrm{alt}}
=
\sum_{v \in \mathcal{V}_{i,K}^{\mathrm{alt}}}
\pi_\theta(v \mid x,y_{<i}).
\label{eq:contrastive-defination}
\end{equation}

We then the contrastive log-odds between the forget target and the alternative can be defined as
\begin{equation}
r_i
=
\log \frac{\pi_i^{\mathrm{target}}}{\pi_i^{\mathrm{alt}}}
=
z_{i,y_i}
-
\log \sum_{v \in \mathcal{V}_{i,K}^{\mathrm{alt}}}
\exp(z_{i,v}),
\label{eq:contrastive-log-odds}
\end{equation}

where $z_{i,v}$ denotes the logit of token $v$ at position $i$.
The second equality follows because target and alternative probabilities share the same softmax normalizer. Thus, $r_i$ measures the relative preference against the selected alternatives as a group.
$r_i>0$ indicates that the target probability exceeds the aggregate probability of the selected alternatives, whereas $r_i<0$ indicates the reverse.
Given a fixed candidate selection, $r_i$ depends only on the target logit and the logits of the selected alternatives.

Let $\beta_i>0$ denote a scaling coefficient that modulates forgetting strength for token $i$, the corresponding loss is $-\log\sigma(-\beta_i r_i)$, where $\sigma$ denotes the sigmoid function. With $\beta_i$ being fixed during differentiation, the loss increases monotonically with $r_i$. Minimizing it therefore reduces the target's relative preference and shifts preference toward plausible non-target alternatives. 
The objective therefore establishes a local competition between the forget target and the selected alternatives, and provides explicit local probability redistribution.

\paragraph{Soft clamping.} 
At the early stage of unlearning, the target probability may substantially exceed the aggregate probability of the selected alternatives, resulting in a large positive $r_i$.
To improve training stability, we apply a tanh-based soft clamping operation \citep{haarnoja2018soft}: 
\begin{equation}
r_i^{clamp}
=
C \tanh\left(\frac{r_i}{C}\right)
\label{eq:soft-clamping}
\end{equation}
where $C>0$ controls the clamping range.
This operation smoothly bounds the contrastive log-odds within $(-C,C)$, preventing extreme margins from dominating the forgetting objective and stabilizing optimization.

\subsection{Adaptive Scaling based on Target Confidence and Partial Entropy}
\label{sec:adaptive_scaling}

While the contrastive log-odds objective establishes token-level competition between the forget target and selected alternatives, a fixed scaling coefficient can not explicitly account for differences in either target confidence or the next-token distribution. 
Target probability may be already low while the remaining probability mass is still concentrated on a small set of non-target tokens. We therefore adapt the coefficient for each token based on both the target confidence and next-token partial entropy. The former captures the model's current preference for the forget target, while the latter characterizes the concentration of the next-token distribution.

\textbf{Target Confidence.}
We use the current probability of the forget target to quantify how strongly the model still favors it under the current token distribution, with higher probability indicates stronger current preference. We define the confidence factor as
\begin{equation}
\label{eq:confidence_factor}
w_i^{\mathrm{conf}}
=
\left(\pi_\theta(y_i \mid x,y_{<i})\right)^\gamma.
\end{equation}

Therefore, stubborn memorized tokens will be assigned with a large factor.
Inspired by focal modulation \citep{lin2017focal}, we further use $\gamma \geq 1$ to control the relative differentiation across tokens. For $\gamma>1$, the confidence factor increases with the target probability. Larger values of $\gamma$ increase the relative emphasis on targets with higher confidence compared with those with lower confidence. When $\gamma=0$, the confidence factor is constant. This factor reflects current target preference rather than directly measuring memorization strength.

\textbf{Next-Token Partial Entropy.}
Target confidence alone does not characterize how probability mass is distributed across the next-token distribution. We therefore complement it with a entropy-based metric computed from the current next-token distribution. Let $\mathcal{T}_{i,K}$ denote the set of $K$ tokens with the highest probabilities under $\pi_\theta(\cdot \mid x,y_{<i})$. Unlike $V_i, k^{alt}$ defined in Section~\ref{para:4.1}, $T_i, K$ does not explicitly exclude the target token. We then define  the next-tokein partial entropy as
\begin{equation}
\label{eq:H_i}
\mathcal{H}_i
=
-\sum_{v \in \mathcal{T}_{i,K}}
\pi_\theta(v \mid x,y_{<i})
\log \pi_\theta(v \mid x,y_{<i}).
\end{equation}

The probabilities are not renormalized within $\mathcal{T}_{i,K}$. Thus, $\mathcal{H}_i$ represents the partial contribution  of the selected tokens to the entropy of the full next-token distribution. It captures both the probability mass covered by the selected tokens and its allocation within that set. We use $\mathcal{H}_i$ as a distributional signal for adaptive scaling, rather than a sufficient measure of distributional collapse.
The corresponding entropy factor is defined as
\begin{equation}
\label{eq:entropy_factor}
w_i^{\mathrm{ent}}
=
\exp\left(\lambda
\left(H_{\mathrm{ref}}-\mathcal{H}_i\right)\right),
\end{equation}

where $\lambda>0$ controls the sensitivity to partial entropy and constant $H_{\mathrm{ref}}$ is a fixed reference level. The factor satisfies $w_i^{\mathrm{ent}}=1$ when $\mathcal{H}_i=H_{\mathrm{ref}}$ and increases as $\mathcal{H}_i$ decreases. It therefore introduces distributional information into the scaling mechanism in addition to the target confidence.

\textbf{Adaptive Forgetting Strength.}
We combine the confidence and entropy factors to define the adaptive scaling coefficient for the contrastive objective:
\begin{equation}
\label{eq:adaptive_strength}
\beta_i
=
\beta_0 \cdot
\operatorname{sg}
\left(w_i^{\mathrm{conf}} w_i^{\mathrm{ent}}\right),
\end{equation}
where $\beta_0>0$ is the base scaling coefficient and $\operatorname{sg}(\cdot)$ denotes the stop-gradient operator.
The factors are computed from the current next-token distribution during each forward pass and treated as constants during differentiation.

The resulting $\beta_i$ depends jointly on target confidence and the partial entropy. For a fixed target confidence, lower partial entropy yields a larger $\beta_i$, so the scaling is not determined by target probability alone. Note that $\beta_i$ modulates the contrastive objective, while the actual gradient magnitude also depends on the sigmoid response and soft clamping.

\subsection{Regime Analysis of RADNPO}
Combining the soft-clamped contrastive log-odds objective defined in (\ref{eq:soft-clamping}) and the token-level adaptive forgetting strength defined in (\ref{eq:adaptive_strength}), we define the forget loss of RADNPO as
\begin{equation}
\mathcal{L}_{\mathrm{RADNPO}} = -\mathbb{E}_{(x,y)\sim\mathcal{D}_f}\left[ \sum_{i=1}^{|y|} \log\sigma\left(-\beta_i r_i^{clamp}\right) \right],
\end{equation}
where $|y|$ is the length of  response sequence. The algorithm of RADNPO is detailed in Appendix \ref{sec:Algorithm}.

In the following, we characterize the token-level unlearning state of each forget target in a two dimensional space defined by the target token probability and next-token partial entropy introduced previously. For notational simplicity, we write $\pi_{\theta,i}$ for $\pi_\theta(y_i\mid x,y_{<i})$.
Importantly, $\beta_i$ is an adaptive coefficient of the local contrastive objective rather than the effective gradient magnitude itself. The latter also depends on the sigmoid response and the derivative of the soft-clamping operation, as detailed in Appendix~\ref{sec:theory}.

\paragraph{Target dominated regime} $(\pi_{\theta,i}\rightarrow 1,\; \mathcal{H}_i\rightarrow 0)$. In this regime, the forget target remains highly preferred and the next-token distribution is strongly concentrated. Consequently, both $w_i^{\mathrm{conf}}$ and $w_i^{\mathrm{ent}}$ take relatively large values, resulting in a larger adaptive coefficient $\beta_i$. RADNPO therefore assigns greater relative scaling to such target-dominated states.

\paragraph{Dispersed forgetting regime} $(\pi_{\theta,i}\rightarrow 0,\; \mathcal{H}_i\text{ relatively high})$.
In this regime, the forget target has a low probability, while probability mass is distributed across multiple non-target tokens and the partial entropy is relatively high. For $\gamma>0$, the low target probability yields a small confidence factor $w_i^{\mathrm{conf}}$. At the same target confidence, a higher partial entropy yields a smaller entropy factor $w_i^{\mathrm{ent}}$ and hence a smaller adaptive coefficient $\beta_i$.

\paragraph{Collapsed forgetting regime} $(\pi_{\theta,i}\rightarrow 0,\; \mathcal{H}_i\rightarrow 0)$.
In this regime, the forget target probability is already low, while the remaining local distribution is still highly concentrated on a small subset of non-target tokens, potentially producing repetitive or uninformative generations. If adaptive scaling depends primarily on the target probability, the optimization coefficient can diminish as $\pi_{\theta,i}$ decreases even though the local distribution remains concentrated.

\begin{figure}[H]
  \centering
  \includegraphics[width=1\linewidth]{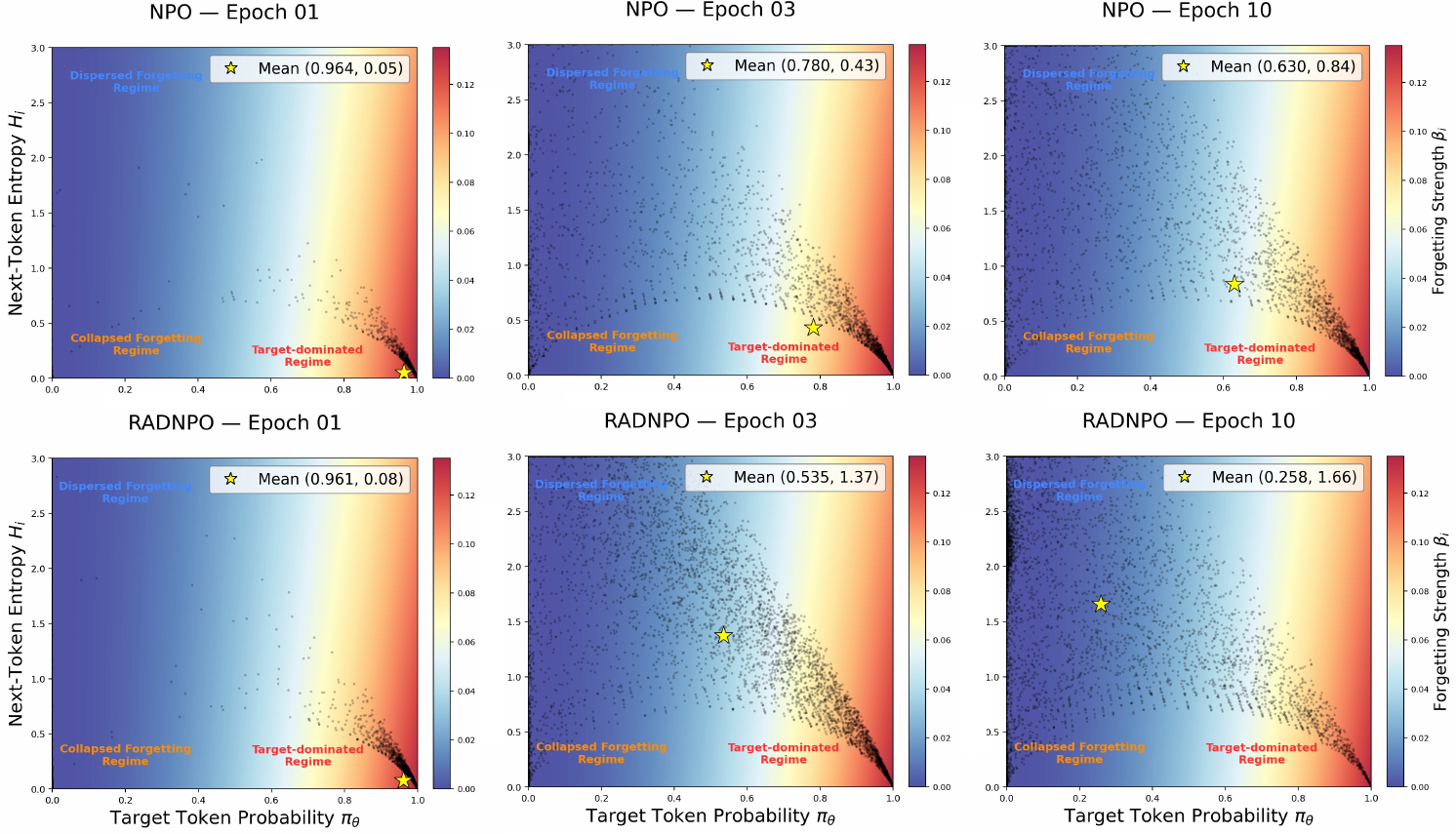}
  \caption{\textbf{Evolution of token states for NPO and RADNPO during unlearning.}The state space is defined by target token probability $\pi_{\theta,i}$ (x-axis) and next-token partial entropy $\mathcal{H}_i$ (y-axis). Background colors visualize the scaling values for each method. Each scatter point represents a token state, and yellow stars mark the mean target probability and mean partial entropy at each displayed epoch. \textbf{NPO:} At later epochs, some token states exhibit both low target probability and low partial entropy. \textbf{RADNPO:} At the same target confidence, lower partial entropy yields a larger adaptive coefficient $\beta_i$. In the later displayed epochs, RADNPO exhibits lower mean target probability and higher mean partial entropy than NPO.
    }
  \label{fig:state_space}
\end{figure}

RADNPO incorporates the partial entropy signal so that a lower $\mathcal{H}_i$ yields a larger $w_i^{\mathrm{ent}}$. At the same target confidence, $w_i^{\mathrm{conf}}$ is identical across states, so a larger entropy factor produces a larger adaptive coefficient $\beta_i$. For the representative regimes described above, this assigns greater relative scaling to the collapsed forgetting regime than to the dispersed forgetting regime. Thus, even after target suppression, the coefficient remains responsive to differences in local distributional statistics rather than depending on target probability alone. Figure~\ref{fig:state_space} illustrates the evolution of target probability and partial entropy during training.

\section{Experiment}
\label{sec:experiments}

\begin{table}[H]
  \centering

  \caption{Unlearning performance on TOFU Forget10 using the LLaMA2-7B-chat model. Best results among the unlearning algorithms are in \textbf{bold} and second best results are \underline{underlined}. Retraining model and original model performances are provided as references. ($\uparrow$) indicates larger values are better, ($\downarrow$) indicates smaller values are better, and ($\rightarrow x$) indicates values closer to $x$ are optimal.}
  \label{tab:main_table}
  \renewcommand{\arraystretch}{1.2}
  \resizebox{\textwidth}{!}{
  \begin{tabular}{l | cccc | c | ccc | c}
    \Xhline{1pt}
    
    \multirow{2}{*}{\textbf{Method}} & 
    \multicolumn{5}{c|}{\textbf{Unlearning Metrics}} & 
    \multicolumn{4}{c}{\textbf{Retain Metrics}} \\
    
    \cline{2-10}
    
    & \makecell{EM \\ $\mathcal{D}_f(\downarrow)$} & \makecell{ES \\ $\mathcal{D}_f(\downarrow)$} & \makecell{Forget TR \\ $\mathcal{D}_f(\downarrow)$} & \makecell{PrivLeak \\ $\mathcal{D}_f(\rightarrow 0)$} & \makecell{\textbf{FQ} \\ $\mathcal{D}_f(\downarrow)$} & \makecell{RA TR \\ $\mathcal{D}_r(\uparrow)$} & \makecell{Retain TR \\ $\mathcal{D}_r(\uparrow)$} & \makecell{WF TR \\ $\mathcal{D}_r(\uparrow)$} & \makecell{\textbf{MU} \\ $\mathcal{D}_r(\uparrow)$} \\
    
    \hline
    
    Original & 0.998 & 0.982 & 0.659 & -99.8 & 56.1 & 0.613 & 0.662 & 0.554 & 0.658 \\
    Retrain & 0.665 & 0.070 & 0.554 & 22.6 & 0.00 & 0.584 & 0.662 & 0.532 & 0.642 \\
    
    \hline
    
    GradAscent & 0.000 & 0.027 & 0.000 & -11.9 & 230 & 1.000 & 0.000 & 1.000 & 0.000 \\
    GradDiff & 0.018 & 0.027 & 0.007 & 59.2 & 230 & 0.769 & 0.585 & 0.626 & 0.061 \\
    
    \hline
    
    WGA & 0.432 & 0.054 & 0.555 & \textbf{3.74} & \underline{3.11} & 0.639 & 0.651 & 0.5997 & 0.664 \\
    RMU & \underline{0.404} & \underline{0.044} & \textbf{0.537} & 37.3 & 5.07 & 0.664 & 0.645 & 0.564 & 0.620 \\
    UNDIAL & 0.621 & 0.056 & 0.606 & -78.1 & 29.3 & \textbf{0.741} & 0.642 & \textbf{0.691} & \underline{0.666} \\
    SatImp & 0.991 & 0.905 & 0.654 & -99.7 & 50.8 & 0.587 &  \underline{0.656} & 0.542 & 0.645 \\
    DPO & 0.916 & 0.452 & 0.620 & -98.5 & 37.7 & 0.592 & 0.629 & 0.525 & 0.498 \\
    NPO & 0.711 & 0.101 & 0.582 & -31.1 & 7.73 & 0.520 & 0.636 & 0.524 & 0.517 \\
    SimNPO & 0.817 & 0.130 & 0.608 & -80.2 & 17.5 & 0.627 & 0.640 & 0.564 & 0.609 \\
    AltPO & 0.607 & 0.091 & 0.578 & -43.2 & 7.52 & 0.663 & 0.638 & 0.543 & 0.649 \\
    
    \hline
    
    \textbf{RADNPO (Ours)} & \textbf{0.382} & \textbf{0.040} & \underline{0.555} & \underline{27.7} & \textbf{0.878} & \underline{0.698} & \textbf{0.658} & \underline{0.658} & \textbf{0.704} \\
    
    \Xhline{1pt}
  \end{tabular}
  }

\end{table}
\subsection{Experiment setups}

\textbf{Datasets and models.}
We evaluate on TOFU \citep{maini2024tofu} and MUSE \citep{shi2024muse} using the Open-Unlearning \citep{dorna2025openunlearning} evaluation framework. 
Furthermore, we leverage Open-unlearning to apply a robust evaluation framework to the  prior benchmarks. 
To ensure consistency with prior work, we adopted LLaMA-2 7B 
for our core evaluations. Furthermore, to demonstrate that our approach generalizes to more recent models, we extended our experiments to Qwen3, one of the latest open-source LLMs, shown in the Appendix \ref{sec:qwen3_exp} .

\textbf{Methods}
We evaluate a range of unlearning methods, including Retrain, GradAscent, GradDiff, DPO \citep{rafailov2023direct}, SimNPO \citep{fan2024simplicity}, NPO \citep{zhang2024negative}, UNDIAL \citep{dong2025undial}, RMU \citep{li2024wmdp}, 
WGA \citep{wang2025rethinking}, SatImp \citep{yang2025exploring}, AltPO\citep{mekala2025alternate}. See Appendix \ref{sec:baseline_method} for more details about the baseline methods.

\subsection{Experiment results}

We include gradient ascent and gradient difference for completeness but exclude them from the main comparison because they cause catastrophic forgetting and severe utility degradation.

\textbf{Performance on TOFU.} 
Table~\ref{tab:main_table} reports the unlearning and retention performance of RADNPO and baselines on TOFU Forget10~\citep{maini2024tofu}. We evaluate five metrics: \textbf{Model Utility (MU)} for retained capability, \textbf{Forget Quality (FQ)} for overall unlearning quality, \textbf{Exact Memorization (EM)} and \textbf{Extraction Strength (ES)}~\citep{dorna2025openunlearning} for forget-set memorization, and \textbf{Privacy Leakage (PrivLeak)} for membership inference risk. Detailed definitions are provided in Appendix~\ref{sec:baseline_eval}.

Table~\ref{tab:main_table} shows that RADNPO removes residual target memorization more effectively than existing PO-based baselines. DPO, NPO, and SimNPO retain relatively high EM and ES, while RADNPO substantially reduces both metrics and achieves an FQ close to the retrain model.
Meanwhile, RADNPO preserves competitive retain performance and achieves the best FQ among the compared methods. These results support our claim that state-aware redistribution improves the forgetting--utility trade-off by suppressing stubborn memories without substantially degrading general capability.

\begin{table}[H]
\centering
\caption{Unlearning performance on MUSE News (LLaMA2-7B) and MUSE Books (ICLM-7B). Best results among the unlearning algorithms are in \textbf{bold} and second best results are \underline{underlined}.}

\label{tab:muse_results}
\resizebox{\textwidth}{!}{
\begin{tabular}{l | cccc | c || cccc | c}
\toprule
\multirow{3}{*}{\textbf{Method}} & \multicolumn{5}{c||}{\textbf{MUSE News}} & \multicolumn{5}{c}{\textbf{MUSE Books}} \\
\cline{2-11}
\rule{0pt}{2ex} & \multicolumn{4}{c|}{\textbf{Unlearning Metrics}} & \textbf{Utility} & \multicolumn{4}{c|}{\textbf{Unlearning Metrics}} & \textbf{Utility} \\
\cline{2-11}
\rule{0pt}{2ex} & EM & ES & VerbMem & KnowMem & KnowMem & EM & ES & VerbMem & KnowMem & KnowMem \\
& $\mathcal{D}_f (\downarrow)$ & $\mathcal{D}_f (\downarrow)$ & $\mathcal{D}_f (\downarrow)$ & $\mathcal{D}_f (\downarrow)$ & $\mathcal{D}_r (\uparrow)$ & $\mathcal{D}_f (\downarrow)$ & $\mathcal{D}_f (\downarrow)$ & $\mathcal{D}_f (\downarrow)$ & $\mathcal{D}_f (\downarrow)$ & $\mathcal{D}_r (\uparrow)$ \\
\midrule
Original        & 0.944 & 0.295 & 0.576 & 0.644 & 0.555 & 0.994 & 0.916 & 0.997 & 0.471 & 0.691 \\
Retrain         & 0.621 & 0.021 & 0.188 & 0.572 & 0.342 & 0.587 & 0.016 & 0.161 & 0.284 & 0.668 \\
\midrule
GradAscent      & 0.163 & 0.009 & 0.006 & 0.008 & 0.121 & 0.157 & 0.008 & 0.008 & 0.007 & 0.100 \\
GradDiff        & 0.194 & 0.007 & 0.011 & 0.029 & 0.213 & 0.201 & 0.008 & 0.014 & 0.031 & 0.198 \\
\midrule
UNDIAL          & \underline{0.726} & \underline{0.031} & \textbf{0.202} & \textbf{0.144} & 0.215 & 0.944 & 0.173 & 0.446 & 0.393 & \underline{0.632} \\
RMU             & 0.903 & 0.098 & 0.389 & 0.542 & \textbf{0.495} & \textbf{0.232} & \underline{0.009} & \underline{0.095} &  \textbf{0.101} & 0.463 \\
WGA             & 0.818 & 0.051 & 0.268 & 0.572 & 0.438 & 0.876 & 0.072 & 0.183 & 0.296 & 0.488 \\
SatImp          & 0.946 & 0.292 & 0.545 & 0.618 & 0.478 & 0.994 & 0.916 & 0.997 & 0.412 & \textbf{0.651} \\
NPO             & 0.788 & 0.038 & \underline{0.204} & 0.467 & 0.390 & 0.916 & 0.129 & 0.294 & \underline{0.236} & 0.537 \\
SimNPO          & 0.941 & 0.243 & 0.527 & 0.615 & \underline{0.488} & 0.906 & 0.180 & 0.302 & 0.308 & 0.506 \\
\midrule
\textbf{RADNPO(Ours)} & \textbf{0.724} & \textbf{0.028} & 0.207 & \underline{0.372} & 0.443 & \underline{0.259} & \textbf{0.008} & \textbf{0.059} & 0.289 & 0.515 \\
\bottomrule
\end{tabular}
}

\end{table}

\textbf{Performance on MUSE. }
Table~\ref{tab:muse_results} reports results on MUSE News and Books. On \textbf{MUSE News}, among utility-preserving baselines, RADNPO achieves one of the strongest forgetting results on MUSE News, indicating stronger removal of memorized content than existing baselines. It also preserves retained knowledge better than aggressive methods such as UNDIAL and improves the forgetting--retention trade-off over NPO.
On \textbf{MUSE Books}, RADNPO remains strong on memorization-related metrics, achieving the lowest ES and lower VerbMem than utility-preserving baselines. Compared with NPO and SimNPO, it forgets more effectively while maintaining competitive retained knowledge. Overall, the results show that RADNPO removes stubborn memories without the severe utility degradation of overly aggressive unlearning methods.

\subsection{Ablation study}

\begin{figure}[H]
  \centering
  \includegraphics[width=\linewidth]{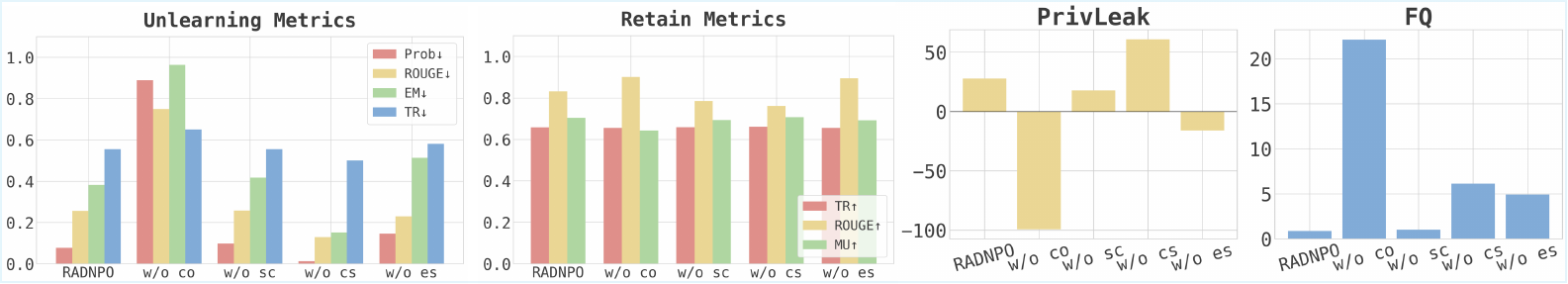}
  \caption{ \textbf{Ablation study of RADNPO.} 
  w/o co stands for without contrastive odds, w/o sc for without soft clamp, w/o cs for without confidence scale, and w/o es for without entropy scale.}
  \label{fig:ablation}

\end{figure}

To assess the contribution of each component, we ablate the contrastive objective, soft clamping, confidence scaling, and entropy scaling. As shown in Figure~\ref{fig:ablation}, removing the contrastive objective causes the largest degradation in both forgetting quality and model utility, indicating that the explicit target-to-alternative comparison is central to the forget--retain trade-off. Removing soft clamping results in a smaller but consistent decline, suggesting that bounding extreme log-ratios improves optimization stability.

The two adaptive scaling factors play complementary roles. Without confidence scaling, forgetting becomes more aggressive while model utility deteriorates, indicating less selective suppression across tokens. Removing entropy scaling primarily reduces forgetting quality with a relatively modest effect on retained utility, suggesting that concentration-aware rescaling helps maintain optimization pressure in low-entropy regimes. Overall, the contrastive objective provides the main redistribution signal, while soft clamping and adaptive scaling improve optimization stability and token-level selectivity.

\subsection{Additional Evaluation and Analysis}

\paragraph{Robustness and sensitivity.}
We further examine RADNPO across different hyperparameter settings and random seeds. The results show relatively stable utility and forget--retain behavior over the evaluated $\lambda_{\mathrm{ent}}$ and $\beta_{\mathrm{base}}$ ranges, while $\gamma_{\mathrm{focal}}$ mainly controls the forgetting operating point. Multi-seed experiments show stable MU, TR, EM, and ES, while FQ and
PrivLeak exhibit larger variation. Details are provided in Appendix~\ref{app:hyperparameter_sensitivity} and Appendix~\ref{app:seed_robustness}.

\paragraph{Generation and semantic quality.}
Beyond usual unlearning metrics, we evaluate generation degeneration and semantic behavior. RADNPO remains substantially closer to the Original model than NPO in repetition and diversity statistics, while the GPT-4o evaluation shows lower forget-set leakage and better retain-side semantic performance. These results provide complementary evidence that RADNPO improves forgetting without severe generation degradation. See Appendix~\ref{app:semantic_eval}.

\paragraph{Representation and recoverability.}
Layer-wise probing shows that RADNPO shifts toward Retrain-like representations earlier than NPO in deeper layers, particularly at layers 30--31, suggesting changes beyond final-layer output suppression. However, fine-tuning the unlearned model on the original forget set substantially restores the forgotten behavior. RADNPO therefore induces representation-level changes but should be interpreted as approximate and recoverable unlearning rather than irreversible erasure. Details are provided in Appendix~\ref{app:representation_probe} and
Appendix~\ref{app:recovery}.

\section{Conclusion}

We analyze a failure mode of LLM unlearning in which substantial target suppression can coexist with highly concentrated local distributions and degraded generation. We propose RADNPO, a reference-free method that combines a local contrastive log-odds objective with adaptive scaling. The objective establishes explicit competition between each forget target and alternative tokens selected from the current next-token distribution. The scaling mechanism adjusts the objective coefficient for each token based on target confidence and partial entropy. Experiments on TOFU and MUSE show favorable trade-offs between forgetting and model utility, while ablations support the contributions of the proposed components. On TOFU Forget10, RADNPO also produces less repetitive and more lexically diverse responses than NPO. Together, these findings support guiding local probability redistribution according to the current token state, rather than relying on target suppression alone.

\section*{AI Use Statement}

During manuscript revision, we used generative AI tools to improve language, clarity, and organization, and to obtain feedback on mathematical explanations, experimental methodology, and the interpretation of results. These tools also assisted in exploring figure layouts. We used the Codex agent to assist with the execution of supplementary experiments. In addition, GPT-4o served as an external judge for the semantic evaluation described in Appendix~\ref{app:semantic_eval}. The authors take full responsibility for the final manuscript, experimental procedures, reported results, and conclusions, including all content produced with AI assistance.

\section*{Ethics Statement}

This work studies approximate unlearning in large language models using the TOFU and MUSE benchmarks. Our intended application is to reduce the reproduction of designated information while preserving useful model behavior. We evaluate forgetting together with retained utility and generation quality, as reducing target memorization alone does not establish reliable information removal. Our results should not be interpreted as a guarantee of complete or irreversible deletion. In particular, the recovery experiments show that forgotten behavior can reappear after fine-tuning on the original forget set. Practical use should therefore include application-specific assessments of residual information leakage and unintended changes to model behavior. The use and redistribution of benchmark data and pretrained models should follow their respective licenses and access conditions.

\section*{Reproducibility}

Section~\ref{sec:method} describes the RADNPO objective and adaptive scaling mechanism, while Appendix~\ref{sec:Algorithm} provides the algorithm. Appendices~\ref{sec:theory} and~\ref{sec:appendix_npo_grad} present the gradient derivations for RADNPO and NPO, respectively. Section~\ref{sec:experiments} and Appendix~\ref{sec:setup} describe the experimental setups, computing resources, and hyperparameter settings. Appendix~\ref{sec:baseline_eval} documents evaluation metrics, and Appendix~\ref{sec:Analysis} describes the additional generation, semantic, representation, and recovery evaluations.

\clearpage

\bibliography{ref}
\bibliographystyle{plainnat}

\clearpage

\section*{Appendix}

\appendix

\section{Algorithm of RADNPO }
\label{sec:Algorithm}
\begin{algorithm}[htbp]
\caption{RADNPO: Reference-free Adaptive Negative Preference Optimization}
\begin{algorithmic}[1]
\State \textbf{Input:} Forget set $\mathcal{D}_f$, retain set $\mathcal{D}_r$; initial model $\theta_{\text{org}}$; learning rate $\eta$
\State \textbf{Parameters:}   $\beta_{0}$, focal factor $\gamma$, entropy factor $\lambda$, ref entropy $H_{ref}$, cutoff $K$, clamp limit $C$
\State \textbf{Output:} Unlearned model $\theta_{\text{radnpo}}$
\State $\theta \leftarrow \theta_{\text{org}}$ \hfill $\triangleright$ Initialize policy
\For{$t = 1$ to $T$}
    \State Sample forget batch $\{(x_f, y_f)\} \sim \mathcal{D}_f$ and retain batch $\{(x_r, y_r)\} \sim \mathcal{D}_r$
    \State \hfill $\triangleright$ \textbf{Stage I: Forget Optimization}
    \For{each token $y_{i}$ in $y_f$}
        \State \hfill $\triangleright$ \textit{1. Soft-clamped Contrastive Odds}
        \State $\pi^{{target}}_i \leftarrow \pi_{\theta}(y_i | x_f, y_{<i})$
        \State $\pi_i^{{alt}} \leftarrow \sum_{v \in \mathcal{V}_{i,K}^{{alt}}}\pi_\theta(v \mid x, y_{<i})$
        \State $r_i \leftarrow \log \frac{\pi^{{target}}_i}{\pi^{{alt}}_i} $
        \State $r_{i}^{clamp} \leftarrow C \cdot \tanh\left(\frac{r_i}{C}\right)$
        \State \hfill $\triangleright$ \textit{2. Dual-State Aware Dynamic Beta}
        \State $w_i^{\mathrm{conf}} \leftarrow \left(\pi_\theta(y_i\mid x,y_{<i})\right)^\gamma$
        \State $\mathcal{T}_{i,K} \leftarrow \operatorname{TopK}_{v} \pi_\theta(v \mid x, y_{<i})$
        \State $\mathcal{H}_i \leftarrow -\sum_{v \in \mathcal{T}_{i,K}} \pi_\theta(v \mid x, y_{<i}) \log \pi_\theta(v \mid x, y_{<i})$
        \State $w_i^{\mathrm{ent}} \leftarrow \exp \left( \lambda \left( \mathcal{H}_{ref} - \mathcal{H}_i \right) \right)$
        \State $\beta_{i} \leftarrow \beta_0 \cdot sg(w_i^{\mathrm{conf}}) \cdot sg(w_i^{\mathrm{ent}})$
        \State \hfill $\triangleright$ \textit{3. Token Forget Loss}
        \State $\mathcal{L}_{\text{forget}}^{(i)} \leftarrow -\log \sigma(-\beta_i \cdot r_{i}^{clamp})$
    \EndFor
    \State $\mathcal{L}_{\text{forget}} \leftarrow  \sum_{i} \mathcal{L}_{\text{forget}}^{(i)}$
    
    \State \hfill $\triangleright$ \textbf{Stage II: Retain Optimization}
    \State Compute normal NLL retain loss $\mathcal{L}_{\text{retain}}$ on $\mathcal{D}_r$
    \State Update model: $\theta \leftarrow \theta - \eta \nabla_{\theta} ( \mathcal{L}_{\text{forget}} +  \mathcal{L}_{\text{retain}})$
\EndFor
\State \textbf{return} $\theta$
\end{algorithmic}
\end{algorithm}

\section{Theoretical analysis of RADNPO gradient rescaling }
\label{sec:theory}

The adaptive coefficient is defined consistently with
Section~\ref{sec:adaptive_scaling} as
\[
\beta_i
=
\beta_0\operatorname{sg}\left[
\pi_\theta(y_i\mid x,y_{<i})^\gamma
\exp\left(\lambda(H_{\mathrm{ref}}-\mathcal{H}_i)\right)
\right].
\]
The stop-gradient operator preserves the forward value but
treats the coefficient as constant during backpropagation.

For a single response $(x,y)$, define
\[
\ell_{\mathrm{RADNPO}}(x,y;\theta)
=
-\sum_{i=1}^{|y|}
\log\sigma\left(-\beta_i r_i^{clamp}\right).
\]
The dataset objective is
\[
\mathcal{L}_{\mathrm{RADNPO}}
=
\mathbb{E}_{(x,y)\sim\mathcal{D}_f}
\left[\ell_{\mathrm{RADNPO}}(x,y;\theta)\right].
\]

With $\beta_i$ held fixed during differentiation,
\[
\nabla_\theta\ell_{\mathrm{RADNPO}}
=
\sum_{i=1}^{|y|}
\beta_i
\sigma\left(\beta_i r_i^{clamp}\right)
\left[1-\tanh^2\left(\frac{r_i}{C}\right)\right]
\nabla_\theta r_i.
\]

\section{Gradient Derivation of the NPO Objective}
\label{sec:appendix_npo_grad}

In this section, we provide the detailed derivation of the NPO gradient
presented in Eq.~\ref{eq:npo_gradient}. The NPO objective can be written as
\begin{equation}
\mathcal{L}_{\mathrm{NPO},\beta}(\theta)
=
\frac{2}{\beta}
\mathbb{E}_{(x,y)\sim\mathcal{D}_{f}}
\left[
\log
\left(
1+
\left(
\frac{\pi_\theta(y|x)}
{\pi_{\mathrm{ref}}(y|x)}
\right)^\beta
\right)
\right].
\label{eq:npo_appendix}
\end{equation}

Let
\begin{equation}
r_\theta(x,y)
=
\log
\frac{\pi_\theta(y|x)}
{\pi_{\mathrm{ref}}(y|x)}.
\end{equation}
Then Eq.~\ref{eq:npo_appendix} can be equivalently written as
\begin{equation}
\mathcal{L}_{\mathrm{NPO},\beta}(\theta)
=
\frac{2}{\beta}
\mathbb{E}_{\mathcal{D}_{f}}
\left[
\log\left(1+\exp(\beta r_\theta)\right)
\right].
\end{equation}

Taking the gradient with respect to $\theta$, we obtain
\begin{equation}
\begin{aligned}
\nabla_{\theta}\mathcal{L}_{\mathrm{NPO},\beta}(\theta)
&=
\frac{2}{\beta}
\mathbb{E}_{\mathcal{D}_{f}}
\left[
\frac{\exp(\beta r_\theta)}
{1+\exp(\beta r_\theta)}
\cdot
\beta\nabla_{\theta}r_\theta
\right]
\\
&=
2\mathbb{E}_{\mathcal{D}_{f}}
\left[
\sigma(\beta r_\theta)
\nabla_{\theta}\log\pi_\theta(y|x)
\right]
\\
&=
2\mathbb{E}_{\mathcal{D}_{f}}
\left[
\frac{
\left(
\frac{\pi_\theta(y|x)}
{\pi_{\mathrm{ref}}(y|x)}
\right)^\beta
}{
1+
\left(
\frac{\pi_\theta(y|x)}
{\pi_{\mathrm{ref}}(y|x)}
\right)^\beta
}
\nabla_{\theta}\log\pi_\theta(y|x)
\right]
\\
&=
\mathbb{E}_{\mathcal{D}_{f}}
\left[
\frac{
2\pi_\theta(y|x)^\beta
}{
\pi_\theta(y|x)^\beta
+
\pi_{\mathrm{ref}}(y|x)^\beta
}
\nabla_{\theta}\log\pi_\theta(y|x)
\right].
\end{aligned}
\label{eq:npo_gradient_derivation}
\end{equation}

Therefore, the effective adaptive weight in NPO is
\begin{equation}
w_{\mathrm{NPO}}(x,y)
=
\frac{
2\pi_\theta(y|x)^\beta
}{
\pi_\theta(y|x)^\beta
+
\pi_{\mathrm{ref}}(y|x)^\beta
}.
\end{equation}

\section{Evaluation Metrics}
\label{sec:baseline_eval}
To comprehensively evaluate the performance of our proposed method, we categorize the evaluation metrics into three distinct groups: Memorization Metrics, Privacy Metrics, and Utility Metrics.

\subsection{Memorization Metrics}
These metrics quantify the extent to which the model has successfully forgotten the targeted information and how much it still memorizes from its training data.

\textbf{Probability (Prob.)}: Directly quantifies the model's confidence in its output on the targeted data. It is computed based on the average loss:
$$ P = \exp \left( -\frac{1}{L} \sum_{t=1}^L -\log P(y_t \mid x, y_{<t}) \right) = \exp(-\text{avg\_loss}) $$

\textbf{ROUGE}: Assesses the degree of overlap between the model's output and the ground truth reference. Specifically, we focus on the ROUGE-$N$-Recall score:
$$ \text{ROUGE-}N\text{-Recall} = \frac{\sum_{\text{n-gram} \in \text{Reference}} \text{Count}_{\text{match}}(\text{n-gram})}{\sum_{\text{n-gram} \in \text{Reference}} \text{Count}(\text{n-gram})} $$

\textbf{Exact Memorization (EM)}: Quantifies memorization by calculating the proportion of tokens in the model's response that exactly match those in the ground truth $y$. Formally, it is defined as:
$$ \text{EM} = \frac{1}{|y|} \sum_k \mathbf{1} \left\{ \arg\max_y f(y \mid [x, y^{<k}]; \boldsymbol{\theta}) = y^k \right\} $$

\textbf{Extraction Strength (ES)}: Quantifies the intensity of memorization by determining the minimal prefix length required to perfectly reconstruct the remaining suffix of the target sequence.
$$ \text{ES} = 1 - \frac{1}{|y|} \min_k \left\{ k \mid f([x, y^{<k}]; \boldsymbol{\theta}) = y^{>k} \right\} $$

\textbf{Truth Ratio (TR)}: Measures the model's preference for the correct answer over a perturbed (incorrect) alternative by comparing their predicted probabilities. A lower value on the forget set indicates successful unlearning.
$$ \text{Truth Ratio} = \frac{p(y^{\text{para}} \mid x)}{p(y^{\text{para}} \mid x) + p(y^{\text{pert}} \mid x)} $$

\textbf{Verbatim Memorization (VerbMem).}
Let $\mathcal{E}_{\mathrm{text}}$ contain evaluation pairs $(c,s)$, where $c$ is a text prefix and $s$ is its reference continuation. We measure verbatim memorization as
\[
\mathrm{VerbMem}
=
\frac{1}{|\mathcal{E}_{\mathrm{text}}|}
\sum_{(c,s)\in\mathcal{E}_{\mathrm{text}}}
\mathrm{ROUGE\mbox{-}L}_{F_1}
\left(g_\theta(c),s\right),
\]
where $g_\theta$ denotes generation under the evaluation prompt and decoding configuration.

\textbf{Knowledge Memorization (KnowMem).}
Let $\mathcal{E}_{\mathrm{QA}}$ contain evaluation question-answer pairs $(q,a)$. We define
\[
\mathrm{KnowMem}
=
\frac{1}{|\mathcal{E}_{\mathrm{QA}}|}
\sum_{(q,a)\in\mathcal{E}_{\mathrm{QA}}}
\mathrm{ROUGE\mbox{-}L}_{F_1}
\left(g_\theta(q),a\right).
\]
We report this metric separately on forget and retain evaluation sets. Lower forget-set scores indicate less reproduction of the queried knowledge, whereas higher retain-set scores indicate better knowledge retention.

\subsection{Privacy Metrics}
These metrics ascertain whether sensitive information from the forget set can still be statistically inferred from the model's behavior.

\textbf{Forget Quality (FQ).}
We compare truth-ratio samples from the unlearned model $f_{\mathrm{unlearn}}$ and the retrained reference model $f_{\mathrm{retrain}}$ on the same forget evaluation set.
Let
\[
\mathcal{S}(f)
=
\left\{
\mathrm{TR}(f;x,y):(x,y)\in\mathcal{D}_f
\right\}.
\]
We obtain the p-value of a two-sample Kolmogorov--Smirnov test:
\[
p_{\mathrm{KS}}
=
\operatorname{pvalue}\left[
\operatorname{KS}_{\mathrm{2samp}}
\left(
\mathcal{S}(f_{\mathrm{unlearn}}),
\mathcal{S}(f_{\mathrm{retrain}})
\right)
\right].
\]
For numerical readability, we report
\[
\mathrm{FQ}
=
-\log_{10}\left(\max(p_{\mathrm{KS}},\epsilon_{\mathrm{KS}})\right).
\]
Lower values correspond to larger KS p-values. This score is used as a distributional diagnostic and does not establish equivalence between the two models.

\textbf{Privacy Leakage (PrivLeak)}
We evaluate membership inference risk by comparing attack AUC scores for the unlearned model and the retrained reference model under the same evaluation protocol. Let $A_\theta$ and $A_{\mathrm{ref}}$ denote the raw attack AUC scores. Following the score orientation used by the evaluation implementation, we define
$\widetilde{A}_\theta=1-A_\theta$ and
$\widetilde{A}_{\mathrm{ref}}=1-A_{\mathrm{ref}}$.
The reported percentage score is
\[
\mathrm{PrivLeak}
=
100\,
\frac{
\widetilde{A}_\theta-\widetilde{A}_{\mathrm{ref}}
}{
\widetilde{A}_{\mathrm{ref}}+\epsilon_{\mathrm{auc}}
}.
\]
Values closer to zero indicate closer agreement with the reference attack AUC under this protocol.

\subsection{Utility Metrics}
The goal of unlearning is to remove the influence of the forget set while preserving the model's performance on non-forget data. Utility metrics therefore evaluate whether the unlearned model retains its capabilities on retained and general knowledge.

\textbf{Retain Truth Ratio (Retain TR).} This metric measures the model's preference for the correct answer over a perturbed alternative on the retain set $D_r$. 

\textbf{Real-Author Truth Ratio (RA TR).} This metric evaluates the same truth-ratio criterion on the Real Authors dataset, which is used to assess whether the model preserves knowledge beyond the forget set and maintain utility on related factual author queries.

\textbf{World-Fact Truth Ratio (WF TR).} This metric applies the truth-ratio evaluation to the World Facts dataset, reflecting whether the model retains broader factual knowledge after unlearning.

\textbf{Model Utility (MU).} MU summarizes retained performance after unlearning. Following prior work, it is computed as a harmonic mean of multiple utility-related metrics across different evaluation sets, including the retain set, Real Authors, and World Facts.

\section{Additional Evaluation Analyses}
\label{sec:Analysis}

\label{app:additional_evaluation}

Unless otherwise stated, the analyses in this section are conducted on LLaMA-2-7B with TOFU Forget10, following the same data processing and evaluation pipeline as the main experiments.
\subsection{Generation-Level and Semantic Evaluation}
\label{app:semantic_eval}

Automatic unlearning metrics mainly measure target removal and retained utility, but do not fully characterize the quality of post-unlearning generations. We therefore complement them with generation-level degeneration metrics and an LLM-based semantic evaluation on LLaMA-2-7B with TOFU Forget10.

\paragraph{Generation-level degeneration.}
We evaluate $n$-gram repetition, Distinct-3, average response length, and Self-BLEU on forget-set generations. Lower repetition and Self-BLEU indicate less repetitive generations, while higher Distinct-3 indicates greater lexical diversity.

\begin{table}[H]
    \centering
    \caption{
    Generation-level statistics on TOFU Forget10. Original is included as a reference. Lower Rep-3, Rep-4, and Self-BLEU and higher Distinct-3 indicate less degenerate generation.
    }
    \label{tab:generation_quality}
    \small
    \setlength{\tabcolsep}{6pt}
    \begin{tabular}{@{}lccccc@{}}
        \toprule
        Model
        & Rep-3 $\downarrow$
        & Rep-4 $\downarrow$
        & Distinct-3 $\uparrow$
        & Avg. Len.
        & Self-BLEU $\downarrow$ \\
        \midrule
        Original
        & 0.0020 & 0.0009 & 0.9980 & 40.4 & 0.157 \\
        NPO
        & 0.0187 & 0.0119 & 0.9813 & 65.4 & 0.268 \\
        RADNPO
        & \textbf{0.0022} & \textbf{0.0011}
        & \textbf{0.9978} & 40.5 & \textbf{0.198} \\
        \bottomrule
    \end{tabular}
\end{table}

NPO exhibits substantially stronger generation degeneration, with markedly higher 3-gram and 4-gram repetition, lower Distinct-3, and higher Self-BLEU. In contrast, RADNPO remains close to the Original model across these statistics. These results provide generation-level evidence that RADNPO mitigates the repetitive and low-diversity behavior associated with concentrated post-unlearning distributions.

\paragraph{LLM-based semantic evaluation.}
We further use GPT-4o as an external judge to evaluate semantic behavior on forget and retain queries. On the forget set, the judge assesses whether the response avoids disclosing target knowledge and measures residual leakage; on the retain set, it evaluates semantic correctness. The overall score summarizes the forget- and retain-side assessments.

\begin{table}[H]
    \centering
    \caption{
    LLM-based semantic evaluation on TOFU Forget10. Higher values are better
    for the overall score, pass rates, and correctness, whereas lower values
    are better for leakage-related metrics.
    }
    \label{tab:semantic_judge}
    \small
    \setlength{\tabcolsep}{3.5pt}
    \resizebox{\linewidth}{!}{%
    \begin{tabular}{@{}lcccccc@{}}
        \toprule
        Method
        & Overall $\uparrow$
        & Forget Pass $\uparrow$
        & Mean Leakage $\downarrow$
        & Retain Pass $\uparrow$
        & Mean Correctness $\uparrow$
        & Partial Leakage $\downarrow$ \\
        \midrule
        NPO
        & 0.371 & 0.370 & 1.280 & 0.373 & 1.120 & 289 \\
        RADNPO
        & \textbf{0.680} & \textbf{0.538} & \textbf{0.995}
        & \textbf{0.823} & \textbf{1.750} & \textbf{131} \\
        \bottomrule
    \end{tabular}%
    }
\end{table}

RADNPO achieves a higher overall judge score than NPO, with lower forget-set leakage and substantially better retain-side semantic performance. Together with the generation-level diagnostics, these results indicate that RADNPO improves the forgetting--retention trade-off without introducing the severe generation degeneration observed for NPO. Since the semantic evaluation relies on an external LLM rather than human annotators, it should be regarded as a diagnostic rather than a substitute for comprehensive human evaluation.

\subsection{Representation-Probing Protocol}
\label{app:representation_probe}

Output-level evaluation alone cannot determine whether unlearning changes internal representations or only suppresses target tokens at the final output layer. To examine this distinction, we train a lightweight binary probe at each selected transformer layer to distinguish hidden representations produced by the Original model, which was trained with the forget data, from those produced by the Retrain model, which was trained without them. We then apply each probe to representations from the unlearned models. The reported Original-like score is the probe-assigned probability of the Original class, averaged over forget samples; lower values therefore indicate a greater shift toward the representation distribution of the Retrain model.

\begin{table}[H]
    \centering
    \caption{Original-like probe scores on forget samples at selected layers.
    Lower values indicate representations closer to those of the Retrain model.}
    \label{tab:representation_probe}
    \small
    \setlength{\tabcolsep}{4.0pt}
    \resizebox{\linewidth}{!}{%
    \begin{tabular}{@{}lccccccccc@{}}
        \toprule
        Method & 1 & 5 & 23 & 25 & 27 & 29 & 30 & 31 & 32 \\
        \midrule
        RADNPO
        & 0.9997 & 0.9996 & 0.9774 & 0.9531 & 0.9151
        & 0.8280 & 0.0710 & 0.0660 & 0.0090 \\
        NPO
        & 0.9995 & 0.9999 & 0.9987 & 0.9972 & 0.9922
        & 0.9709 & 0.6550 & 0.4800 & 0.0010 \\
        \bottomrule
    \end{tabular}%
    }
\end{table}

As shown in Table~\ref{tab:representation_probe}, RADNPO departs from Original-like representations earlier in the deepest intermediate layers. The difference is most pronounced at layers 30 and 31, whereas both methods obtain near-zero scores at the final layer. This pattern suggests that RADNPO affects internal representations before the output layer rather than acting only on the final logits. Nevertheless, a probe measures distributional similarity under a specific diagnostic protocol and does not establish irreversible knowledge
erasure.

\subsection{Recovery Evaluation}
\label{app:recovery}

We assess reversibility by initializing from the RADNPO-unlearned checkpoint and applying standard supervised fine-tuning on the original forget set. The recovered model returns close to the Original model on forget-target metrics, showing that RADNPO's behavioral forgetting is substantially recoverable under direct re-exposure to the deleted data. RADNPO should therefore be interpreted as approximate behavioral unlearning rather than certified irreversible
deletion.

\begin{table}[H]
    \centering
    \caption{Recovery evaluation after supervised fine-tuning of the
    RADNPO-unlearned model on the original forget set.}
    \label{tab:recovery}
    \small
    \setlength{\tabcolsep}{3.0pt}
    \resizebox{\linewidth}{!}{%
    \begin{tabular}{@{}lccccccccc@{}}
        \toprule
        Method
        & EM $\downarrow$
        & ES $\downarrow$
        & Forget TR $\downarrow$
        & PrivLeak $\rightarrow 1$
        & FQ $\downarrow$
        & RA TR $\uparrow$
        & Retain TR $\uparrow$
        & WF TR $\uparrow$
        & MU $\uparrow$ \\
        \midrule
        Original
        & 0.998 & 0.982 & 0.659 & -99.8 & 56.1
        & 0.613 & 0.662 & 0.554 & 0.658 \\
        Retrain
        & 0.665 & 0.070 & 0.554 & 22.6 & 0.00
        & 0.584 & 0.662 & 0.532 & 0.642 \\
        RADNPO
        & 0.382 & 0.040 & 0.555 & 27.7 & 0.878
        & 0.698 & 0.658 & 0.658 & 0.704 \\
        Recovered
        & 0.968 & 0.976 & 0.647 & -97.2 & 49.6
        & 0.614 & 0.663 & 0.637 & 0.684 \\
        \bottomrule
    \end{tabular}%
    }
\end{table}

This experiment specifically evaluates recovery after fine-tuning on the exact forget set. It does not imply that arbitrary downstream fine-tuning or incidental exposure to related data will necessarily recover the forgotten behavior.

\section{Baseline Methods}
\label{sec:baseline_method}

\textbf{Retrain}: Serves as the gold standard for machine unlearning by training the model entirely from scratch utilizing only the retain dataset, ensuring complete isolation from the forget data.

\textbf{GradAscent.}
This baseline performs gradient ascent on the negative
log-likelihood of forget responses to suppress their likelihood.
Equivalently, it minimizes the following log-likelihood objective:
\[
\mathcal{L}(\theta)
=
\mathbb{E}_{(x,y)\sim\mathcal{D}_f}
\left[\log\pi_\theta(y\mid x)\right].
\]

\textbf{GradDiff.}
This baseline combines gradient ascent on the negative
log-likelihood of forget responses with gradient descent on
the negative log-likelihood of retain responses.
The resulting objective balances target suppression with
preservation of retained performance:
\[
\begin{aligned}
\mathcal{L}(\theta)
={}&
\mathbb{E}_{(x,y)\sim\mathcal{D}_f}
\left[\log\pi_\theta(y\mid x)\right]
\\
&-
\lambda_{\mathrm{retain}}
\mathbb{E}_{(x,y)\sim\mathcal{D}_r}
\left[\log\pi_\theta(y\mid x)\right],
\end{aligned}
\]
where $\lambda_{\mathrm{retain}}>0$ controls the relative
weight of the retain objective.

\textbf{DPO} \citep{rafailov2023direct}.
We adapt Direct Preference Optimization to unlearning by
pairing each forget response $y_f$ with a designated
``I don't know'' response $y_{\mathrm{idk}}$ for the same
prompt $x$.
The refusal response is treated as preferred and the original
forget response as dispreferred.
The preference objective is
\[
\mathcal{L}(\theta)
=
-\frac{2}{\beta}
\mathbb{E}_{(x,y_f)\sim\mathcal{D}_f}
\left[
\log\sigma\left(
\beta\log
\frac{\pi_\theta(y_{\mathrm{idk}}\mid x)}
     {\pi_{\mathrm{ref}}(y_{\mathrm{idk}}\mid x)}
-
\beta\log
\frac{\pi_\theta(y_f\mid x)}
     {\pi_{\mathrm{ref}}(y_f\mid x)}
\right)
\right],
\]
where $\sigma$ is the sigmoid function, $\beta>0$ controls
the scaling of the preference margin, and $\pi_{\mathrm{ref}}$
is a frozen reference model.
Minimizing this objective favors the refusal response over
the forget response in terms of their relative log-likelihoods
with respect to the reference model.

\textbf{SimNPO} \cite{fan2024simplicity}: A streamlined variant of Negative Preference Optimization (NPO) that eliminates the reliance on a reference model to mitigate reference bias and ensure more balanced optimization across the forget set.
$$\mathcal{L} = \frac{2}{\beta} \mathbb{E}_{(x,y) \sim \mathcal{D}_{f}} \log \sigma \left( - \frac{\beta}{|y|} \log \pi_{\theta}(y|x) - \delta \right)$$

\textbf{AltPO} \cite{mekala2025alternate}: 
Alternate Preference Optimization combines negative feedback on the original
forget response with positive feedback from prompt-specific, in-domain
alternative responses. For each forget pair $(x_f,y_f)$, AltPO generates
alternative labels $y_a$ and applies a DPO-style objective that increases the
relative preference for $y_a$ while suppressing $y_f$, together with an NLL
loss on the retain set to preserve model utility. Its objective is

$$
\mathcal{L}
=
\mathbb{E}_{(x,y)\sim\mathcal D_f,\;y_a\sim\mathcal A(x)}
[\mathcal L_{\mathrm{DPO}}(y_a,y|x)]
$$

\textbf{UNDIAL} \cite{dong2025undial}: Leverages self-distillation to smoothly adjust output logits toward a uniform distribution for targeted tokens, ensuring stable convergence and mitigating the risk of over-unlearning. The adjusted logits are defined as:
$$z_{\text{adj}}(x) = z_{\text{orig}}(x) - \beta \cdot \mathbf{1}_{y_f}$$
The core idea is achieved by minimizing the KL divergence between the adjusted logits and the model's current output distribution:
$$\mathcal{L} =  \mathbb{E}_{(x,y) \sim \mathcal{D}_{f}} \left[ \text{KL} \left( \text{softmax}(z_{\text{adj}}(x)) \| \text{softmax}(z_{\text{unl}}(x)) \right) \right] + \lambda \mathbb{E}_{(x,y) \sim \mathcal{D}_{r}} \pi_{\theta}(y|x)$$
Where $z_{\text{orig}}(x)$ is the original logits produced by the model before unlearning and $z_{\text{adj}}(x)$ is the adjusted logits.

\textbf{RMU} \cite{li2024wmdp}: Perturbs the model's hidden states to misdirect the internal representations of forget data into a random or irrelevant subspace, effectively rendering the model incoherent on targeted queries while preserving retain performance. Let $\phi(s; f_{\text{unl}})$ denote the embedding features of the model, the loss is given by:
$$\mathcal{L} = \mathbb{E}_{(x,y)\sim \mathcal{D}_f} \frac{1}{|y_f|} \sum_{i=1}^{|y_f|} \|\phi([x, y^{<i}]) - c \cdot u\|_2^2 + \mathbb{E}_{(x,y)\sim \mathcal{D}_r} \frac{1}{|y|} \sum_{i=1}^{|y|} \|\phi([x, y^{<i}]) - \phi_{\text{ref}}([x, y^{<i}])\|_2^2 $$
where $u$ has elements randomly sampled from $[0, 1)$ and $c$ is a scaling hyper-parameter.

\textbf{WGA} \cite{wang2025rethinking}: Enhances vanilla Gradient Ascent by incorporating a confidence-based loss weighting mechanism, which prevents excessive parameter updates and mitigates unnecessary degradation of model integrity.
$$ \mathcal{L} = \mathbb{E}_{(x,y)\sim \mathcal{D}_f} \left[ \left( \exp(-\ell_{\text{CE}}) \right)^{\beta} \cdot \ell_{\text{CE}} \right] $$

\textbf{SatImp} \cite{yang2025exploring}: Employs a dual-criteria loss reweighting strategy that simultaneously targets "Saturation" (insufficiently optimized data) and "Importance" (critical data) to optimize the unlearning trajectory.
$$ \mathcal{L} = \mathbb{E}_{(x,y)\sim \mathcal{D}_f} \left[ \left( \exp(-\ell_{\text{CE}}) \right)^{\beta_1} \cdot \left( 1 - \exp(-\ell_{\text{CE}}) \right)^{\beta_2} \cdot \ell_{\text{CE}} \right] $$
where $\beta_1$ controls the saturation weight and $\beta_2$ controls the importance weight.

\section{Additional Experiment Details and Results}
\label{sec:qwen3}

\subsection{Computing Resources}
\label{sec:resource}
All experiments are conducted on 4 NVIDIA 4090 GPU in a single node.

\subsection{Experiment Setups}
\label{sec:setup}
For our experiments with the proposed RADNPO method, we employ a linear warm-up learning rate during the first epoch, followed by a linearly decaying learning rate in the remaining epochs. We initialize the unlearning process with the LLaMA-2 7B model previously fine-tuned on the TOFU dataset. RADNPO is trained for 10 epochs with an effective batch size of 32 and a peak learning rate of $10^{-5}$, utilizing the AdamW optimizer with a weight decay of 0.01. For the RADNPO-specific hyperparameters on the TOFU benchmark, we set the base dynamic temperature $\beta_{base} = 0.1$, the focal memorization scale exponent $\gamma_{focal} = 1.0$, and the entropy-based temperature scaling factor $\lambda_{ent} = 0.1$. Conversely, for experiments conducted on the MUSE benchmark, we adjust the dynamic parameters to $\beta_{base} = 0.3$ and $\gamma_{focal} = 0.5$, while maintaining $\lambda_{ent} = 0.1$. 
We set $H_{\mathrm{ref}}=10$ as a fixed scaling reference, rather than a theoretical maximum entropy, and use $K=10$ for the top-$K$ partial-entropy computation.
We use $K=10$ for both the alternative candidate set and the partial entropy computation on TOFU and MUSE. The former excludes the forget target, whereas the latter is selected from the full vocabulary. We fix $H_{\mathrm{ref}}=10$ throughout training as a scaling reference rather than a theoretical maximum entropy.
Additionally, log-odds are clipped at 8.0 to prevent gradient explosion. The loss weights for the retain and forget objectives are uniformly set to $\alpha = 1.0$ and $\gamma = 1.0$, respectively, using Negative Log-Likelihood for the retain loss. All other data processing and evaluation pipelines strictly follow the Open-Unlearning \citep{dorna2025openunlearning} benchmark setups.

For the TOFU benchmark, our setup is strictly integrated with the open-unlearning framework. Rather than training from scratch, we directly utilize their publicly available, pre-trained checkpoints---specifically, the models trained on the TOFU FULL split and the TOFU Retain90 split---to initialize our unlearning process and conduct evaluations. In contrast, when extending our unlearning framework to more recent base architectures, namely LLaMA-3-8B and Qwen3, pre-trained original and retrained checkpoints are not directly adopted. Instead, we rigorously follow the standard Supervised Fine-Tuning (SFT) pipeline to generate both the Original and Retrain models from the ground up before applying our unlearning methods.
Furthermore, for evaluations on the MUSE benchmark, we investigate two distinct domains. For MUSE News, we use LLaMA-2 7B fine-tuned on BBC news articles as the original model. For MUSE Books, we use ICLM 7B fine-tuned on the Harry Potter books as the original model. The original models for both Books and News can be directly obtained from the benchmark repositories.

\subsection{Sensitivity to RADNPO Hyperparameters}
\label{app:hyperparameter_sensitivity}

We further examine the sensitivity of RADNPO to its main adaptive-scaling
hyperparameters on LLaMA-2-7B with TOFU Forget10. Specifically, we vary the
focal exponent $\gamma_{\mathrm{focal}} \in \{1,2\}$, the entropy scaling
coefficient $\lambda_{\mathrm{ent}} \in \{0.1,0.2,0.5\}$, and the base
temperature $\beta_{\mathrm{base}} \in \{0.1,0.3\}$, while keeping all other
training settings fixed.

\begin{table}[!ht]
\centering
\caption{
Sensitivity of RADNPO to $\gamma_{\mathrm{focal}}$,
$\lambda_{\mathrm{ent}}$, and $\beta_{\mathrm{base}}$ on TOFU Forget10.
The bold row denotes the default configuration used in the main experiments.
}
\label{tab:radnpo_hyperparameter_sensitivity}

\small
\renewcommand{\arraystretch}{1.08}
\setlength{\tabcolsep}{4.2pt}

\resizebox{\textwidth}{!}{
\begin{tabular}{ccc|ccccccc}
\toprule
$\gamma_{\mathrm{focal}}$
& $\lambda_{\mathrm{ent}}$
& $\beta_{\mathrm{base}}$
& MU $\uparrow$
& FQ $\downarrow$
& Forget TR $\downarrow$
& Retain TR $\uparrow$
& EM $\downarrow$
& ES $\downarrow$
& PrivLeak $\rightarrow 1$ \\
\midrule

\textbf{1.0}
& \textbf{0.1}
& \textbf{0.1}
& \textbf{0.704}
& \textbf{0.878}
& \textbf{0.555}
& \textbf{0.658}
& \textbf{0.382}
& \textbf{0.040}
& \textbf{27.7} \\

1.0 & 0.2 & 0.1
& 0.706 & 0.322 & 0.554 & 0.659 & 0.371 & 0.038 & 29.4 \\

1.0 & 0.5 & 0.1
& 0.706 & 0.416 & 0.555 & 0.658 & 0.383 & 0.039 & 26.6 \\

\midrule

2.0 & 0.1 & 0.1
& 0.701 & 0.054 & 0.566 & 0.658 & 0.466 & 0.046 & 8.16 \\

2.0 & 0.2 & 0.1
& 0.701 & 0.065 & 0.566 & 0.658 & 0.467 & 0.047 & 8.41 \\

2.0 & 0.5 & 0.1
& 0.702 & 0.045 & 0.567 & 0.658 & 0.469 & 0.047 & 7.72 \\

\midrule

1.0 & 0.1 & 0.3
& 0.705 & 0.410 & 0.557 & 0.657 & 0.385 & 0.040 & 27.0 \\

1.0 & 0.2 & 0.3
& 0.707 & 0.320 & 0.556 & 0.658 & 0.374 & 0.038 & 28.7 \\

1.0 & 0.5 & 0.3
& 0.707 & 0.410 & 0.557 & 0.657 & 0.386 & 0.040 & 26.2 \\

\midrule

2.0 & 0.1 & 0.3
& 0.702 & 0.055 & 0.568 & 0.657 & 0.468 & 0.046 & 8.00 \\

2.0 & 0.2 & 0.3
& 0.702 & 0.066 & 0.568 & 0.657 & 0.469 & 0.047 & 8.20 \\

2.0 & 0.5 & 0.3
& 0.703 & 0.046 & 0.569 & 0.657 & 0.471 & 0.047 & 7.60 \\

\bottomrule
\end{tabular}
}
\end{table}

As shown in Table~\ref{tab:radnpo_hyperparameter_sensitivity},
RADNPO is relatively stable with respect to
$\lambda_{\mathrm{ent}}$ and $\beta_{\mathrm{base}}$: varying either
parameter produces only modest changes in model utility, forget/retain truth
ratios, and memorization metrics. In contrast,
$\gamma_{\mathrm{focal}}$ has a more pronounced effect on the forgetting
operating point. Increasing it from 1 to 2 substantially reduces FQ and brings
PrivLeak closer to its desired value, while slightly increasing EM, ES, and
Forget TR. This indicates that the focal exponent primarily controls the
relative emphasis placed on highly confident forget targets, whereas the
method remains comparatively stable over the evaluated entropy-scaling and
base-temperature ranges.

\subsection{Robustness Across Random Seeds}
\label{app:seed_robustness}

The main results follow the fixed-seed evaluation protocol commonly used in
existing LLM unlearning benchmarks. To further examine robustness to training
randomness, we additionally evaluate RADNPO with three random seeds on
LLaMA-2-7B with TOFU Forget10. The results are reported in
Table~\ref{tab:seed_robustness}.

\begin{table}[!ht]
\centering
\caption{
Robustness of RADNPO across different random seeds on TOFU Forget10.
}
\label{tab:seed_robustness}
\small
\renewcommand{\arraystretch}{1.08}
\setlength{\tabcolsep}{5pt}
\begin{tabular}{c|ccccccc}
\toprule
Seed
& MU $\uparrow$
& FQ $\downarrow$
& Forget TR $\downarrow$
& Retain TR $\uparrow$
& EM $\downarrow$
& ES $\downarrow$
& PrivLeak $\rightarrow 1$ \\
\midrule
42
& 0.702 & 0.046 & 0.565 & 0.658 & 0.472 & 0.047 & 7.9 \\
123
& 0.704 & 0.430 & 0.567 & 0.657 & 0.464 & 0.046 & 17.6 \\
2026
& 0.701 & 0.850 & 0.566 & 0.659 & 0.469 & 0.047 & 5.8 \\

\bottomrule
\end{tabular}
\end{table}

Across the three seeds, RADNPO shows highly stable MU, Forget TR, Retain TR, EM, and ES, indicating that its overall forgetting--utility behavior is largely consistent across runs. FQ and PrivLeak exhibit larger numerical variation, suggesting that these distribution- and privacy-based metrics are more sensitive to training randomness. Overall, the multi-seed results support the robustness of the main performance trends, while we do not claim statistical significance from three runs alone.

\subsection{Sensitivity to the Retain-Forget Loss Balance}
\label{app:loss_balance}

We study the sensitivity of RADNPO to the balance between retain and forget losses on LLaMA-2-7B with TOFU Forget10. For this sweep, we use normalized loss weights:
\begin{equation}
\mathcal{L}
=
\lambda_{\mathrm{retain}}\mathcal{L}_{\mathrm{retain}}
+
\lambda_{\mathrm{forget}}\mathcal{L}_{\mathrm{RADNPO}},
\qquad
\lambda_{\mathrm{retain}}+\lambda_{\mathrm{forget}}=1.
\end{equation}
We vary $\lambda_{\mathrm{retain}}$ over $\{0.2,0.4,0.6,0.8\}$ and set
$\lambda_{\mathrm{forget}}=1-\lambda_{\mathrm{retain}}$.
This normalized-weight sweep is separate from the main-experiment
configuration, which uses unit weights for both losses.

\begin{table}[!ht]
\centering
\caption{
Sensitivity to normalized retain--forget loss weights on TOFU Forget10.
Arrows indicate the preferred direction of each metric.
}
\label{tab:loss_balance}
\small
\setlength{\tabcolsep}{6pt}
\renewcommand{\arraystretch}{1.05}
\begin{tabular}{@{}cccccc@{}}
\toprule
$\lambda_{\mathrm{retain}}$
& $\lambda_{\mathrm{forget}}$
& MU $\uparrow$
& FQ $\downarrow$
& Forget TR $\downarrow$
& Retain TR $\uparrow$ \\
\midrule
0.8 & 0.2 & \textbf{0.701} & 4.620
    & 0.592 & \textbf{0.662} \\
0.6 & 0.4 & 0.691 & \textbf{0.098}
    & 0.553 & 0.658 \\
0.4 & 0.6 & 0.692 & 0.131
    & 0.561 & 0.657 \\
0.2 & 0.8 & 0.575 & 1.260
    & \textbf{0.532} & 0.617 \\
\bottomrule
\end{tabular}
\end{table}

Table~\ref{tab:loss_balance} shows that intermediate retain weights ($0.4$ and $0.6$) achieve lower FQ while maintaining comparable utility. The retain-heavy setting ($0.8$) yields the highest MU and Retain TR, but worse FQ. Conversely, the forget-heavy setting ($0.2$) lowers Forget TR at the cost of retained utility. These results highlight the importance of balancing the two objectives, with intermediate weights providing favorable trade-offs in this experiment.

\subsection{Extended experiments on LLaMA3-8B}

\begin{table}[H]
  \centering
  \caption{Unlearning performance on TOFU Forget10 using the LLaMA3-8B-chat model. Retraining model and original model performances are provided as references. ($\uparrow$) indicates larger values are better, ($\downarrow$) indicates smaller values are better, and ($\rightarrow x$) indicates values closer to $x$ are optimal.}
  \label{tab:llama3_table}
  \renewcommand{\arraystretch}{1.2}
  \resizebox{\textwidth}{!}{
  \begin{tabular}{l | cccc | c | ccc | c}
    \Xhline{1pt}
    
    \multirow{2}{*}{\textbf{Method}} & 
    \multicolumn{5}{c|}{\textbf{Unlearning Metrics}} & 
    \multicolumn{4}{c}{\textbf{Retain Metrics}} \\
    
    \cline{2-10}
    
    & \makecell{EM \\ $\mathcal{D}_f(\downarrow)$} & \makecell{ES \\ $\mathcal{D}_f(\downarrow)$} & \makecell{Forget TR \\ $\mathcal{D}_f(\downarrow)$} & \makecell{PrivLeak \\ $\mathcal{D}_f(\rightarrow 0)$} & \makecell{\textbf{FQ}  \\ $\mathcal{D}_f(\downarrow)$} & \makecell{RA TR \\ $\mathcal{D}_r(\uparrow)$} & \makecell{Retain TR \\ $\mathcal{D}_r(\uparrow)$} & \makecell{WF TR \\ $\mathcal{D}_r(\uparrow)$} & \makecell{\textbf{MU} \\ $\mathcal{D}_r(\uparrow)$} \\
    
    \hline
    
    Original  & 0.998 & 0.979 & 0.686 & -99.9 & 26.44 & 0.494 & 0.696 & 0.621 & 0.647 \\
    Retrain & 0.613 & 0.065 & 0.562 & 24.1 & 0.00 & 0.546 & 0.690 & 0.642 & 0.671 \\
    
    \hline
    
    WGA & 0.444 & 0.039 & 0.549 & 42.3 & 6.86 & 0.545 & 0.686 & 0.663 & 0.663 \\
    RMU & 0.231 & 0.040 & 0.532 & 52.1 & 5.04 & 0.581 & 0.693 & 0.685 & 0.695 \\
    SatImp & 0.980 & 0.820 & 0.632 & -99.6 & 20.86 & 0.639 & 0.680 & 0.549 & 0.660 \\
    UNDIAL & 0.761 & 0.121 & 0.649 & -96.5 & 20.47 & 0.734 & 0.667 & 0.779 & 0.708 \\
    DPO & 0.856 & 0.344 & 0.632 & -95.7 & 14.2 & 0.614 & 0.648 & 0.696 & 0.663 \\
    NPO & 0.588 & 0.068 & 0.582 & 4.4 & 3.36 & 0.601 & 0.651 & 0.630 & 0.521 \\
    SimNPO & 0.947 & 0.533 & 0.678 & -99.2 & 24.02 & 0.531 & 0.685 & 0.640 & 0.656 \\
    RADNPO & 0.323 & 0.038 & 0.660 & 4.8 & 3.24 & 0.479 & 0.677 & 0.612 & 0.696 \\
    
    \Xhline{1pt}
  \end{tabular}
  }
\end{table}

\subsection{Extended experiments on Qwen3-4B}
\label{sec:qwen3_exp}
\begin{table}[H]
  \centering
  \caption{Unlearning performance on TOFU Forget10 using the Qwen3-4B-Instruct model. Retraining model and original model performances are provided as references. ($\uparrow$) indicates larger values are better, ($\downarrow$) indicates smaller values are better, and ($\rightarrow x$) indicates values closer to $x$ are optimal.}
  \label{tab:qwen3_table}
  \renewcommand{\arraystretch}{1.2}
  \resizebox{\textwidth}{!}{
  \begin{tabular}{l | cccc | c | ccc | c}
    \Xhline{1pt}
    
    \multirow{2}{*}{\textbf{Method}} & 
    \multicolumn{5}{c|}{\textbf{Unlearning Metrics}} & 
    \multicolumn{4}{c}{\textbf{Retain Metrics}} \\
    
    \cline{2-10}
    
    & \makecell{EM \\ $\mathcal{D}_f(\downarrow)$} & \makecell{ES \\ $\mathcal{D}_f(\downarrow)$} & \makecell{Forget TR \\ $\mathcal{D}_f(\downarrow)$} & \makecell{PrivLeak \\ $\mathcal{D}_f(\rightarrow 0)$} & \makecell{\textbf{FQ}  \\ $\mathcal{D}_f(\downarrow)$} & \makecell{RA TR \\ $\mathcal{D}_r(\uparrow)$} & \makecell{Retain TR \\ $\mathcal{D}_r(\uparrow)$} & \makecell{WF TR \\ $\mathcal{D}_r(\uparrow)$} & \makecell{\textbf{MU} \\ $\mathcal{D}_r(\uparrow)$} \\
    
    \hline
    
    Original  & 0.992 & 0.874 & 0.683 & -99.5 & 17.44 & 0.380 & 0.688 & 0.509 & 0.559 \\
    Retrain & 0.617 & 0.105 & 0.579 & 26.7 & 0.00 & 0.403 & 0.685 & 0.583 & 0.592 \\
    
    \hline
    
    WGA & 0.657 & 0.116 & 0.633 & -70.1 & 7.59 & 0.414 & 0.669 & 0.537 & 0.559 \\
    RMU & 0.728 & 0.150 & 0.641 & -82.6 & 6.69 & 0.370 & 0.664 & 0.492 & 0.467 \\
    SatImp & 0.942 & 0.530 & 0.669 & -99.1 & 14.76 & 0.479 & 0.675 & 0.530 & 0.587 \\
    UNDIAL & 0.749 & 0.138 & 0.643 & -93.8 & 13.24 & 0.558 & 0.661 & 0.622 & 0.601 \\
    DPO & 0.749 & 0.183 & 0.610 & -78.1 & 7.22 & 0.605 & 0.619 & 0.592 & 0.417 \\
    NPO & 0.644 & 0.108 & 0.587 & -16.6 & 3.37 & 0.629 & 0.609 & 0.508 & 0.478 \\
    SimNPO & 0.616 & 0.105 & 0.574 & 6.12 & 3.13 & 0.435 & 0.675 & 0.617 & 0.581 \\
    RADNPO & 0.551 & 0.107 & 0.639 & -11.7 & 2.36 & 0.645 & 0.678 & 0.663 & 0.660 \\
    
    \Xhline{1pt}
  \end{tabular}
  }
\end{table}

\end{document}